\documentclass[prb,aps,twocolumn,groupedaddress,floats,showpacs,final,superscriptaddress]{revtex4-2}
\usepackage[latin3]{inputenc}
\usepackage[makeroom]{cancel}
\usepackage{graphicx}
\usepackage{amsmath}
\usepackage{amsfonts}
\usepackage{amssymb}
\usepackage{chngcntr}
\usepackage{color}
\usepackage[left=2cm,right=2cm,top=2cm,bottom=2cm]{geometry}
\usepackage{graphicx}
\usepackage{dcolumn}
\usepackage{bm}
\usepackage{simplewick}
\usepackage{array}
\usepackage{appendix}

\RequirePackage[
   hyperindex,colorlinks,bookmarksnumbered,
   plainpages=true,pdfstartview=FitH]{hyperref}
\hypersetup{linkcolor=blue,urlcolor=blue,citecolor=blue}
\usepackage{hyperref}

\definecolor{purple}{rgb}{0.5,0,0.6}

\begin{document}



\title{Renormalization of the valley Hall conductivity due to interparticle interaction}


\author{D.~S.~Eliseev}
\affiliation{Novosibirsk State Technical University, Novosibirsk 630073, Russia}

\author{A.~V.~Parafilo}
\affiliation{Center for Theoretical Physics of Complex Systems, Institute for Basic Science (IBS), Daejeon 34126, Korea}

\author{V.~M.~Kovalev}
\affiliation{Novosibirsk State Technical University, Novosibirsk 630073, Russia}

\author{O.~V.~Kibis}
\affiliation{Novosibirsk State Technical University, Novosibirsk 630073, Russia}


\author{I.~G.~Savenko}
\affiliation{Department of Physics, Guangdong Technion--Israel Institute of Technology, 241 Daxue Road, Shantou, Guangdong 15063, China}
\affiliation{Technion -- Israel Institute of Technology, 32000 Haifa, Israel}
\affiliation{Guangdong Provincial Key Laboratory of Materials and Technologies for Energy Conversion, Guangdong Technion--Israel Institute of Technology, Guangdong 515063, China}

\date{\today}

\begin{abstract}
We develop a theory of Coulomb interaction-mediated contribution to valley Hall effect (VHE) in two-dimensional non-centrosymmetric gapped Dirac materials.
We assume that the bare valley Hall current occurs in the system due to the presence of disorder caused by impurities and is determined by the valley-selective anisotropic skew scattering.
Applying the Boltzmann transport equation to describe the electron and hole transport in the material, we calculate the renormalized VHE conductivity due to electron-electron and electron-hole scattering processes, considering two regimes: (i) an $n$-doped monolayer hosting a degenerate electron gas, and (ii) an intrinsic semiconductor with the Boltzmann statistics of electron and hole gases.
In both regimes, the dominant mechanism of interparticle scattering is due to particles residing in different valleys.
Moreover, in case (ii), in addition to direct scattering, electron-hole annihilation starts to play a role with the increase in temperature.
It might even become the dominant mechanism of the Coulomb interaction-mediated VHE.
\end{abstract}

\maketitle


{\it Introduction.}
The Coulomb scattering of the carriers of charge in solids is mostly important at sufficiently low temperatures -- such temperatures at which the scattering on optical and acoustic phonons is ``frozen''.
In semiconductors, electrons and holes can scatter on each other;
thus, electron-electron (e-e), hole-hole (h-h), and electron-hole (e-h) processes take place.
They determine the temperature behavior of the transport coefficients~\cite{Lithuanian}.
The efficiency of each of these processes depends on the density of particles: In an intrinsic semiconductor (the insulating phase), e-h scattering can be dominating~\cite{arXiv:2403.10898};
instead, in the degenerate electron (hole) gas case, when the Fermi level lies in either the conduction or the valence band due to doping, the expected dominant mechanism is e-e or h-h, depending on the position of the Fermi level.
Under optical excitation, Fermi quasi-levels might cross both the valence and conduction bands simultaneously (it corresponds to non-equilibrium), and the properties of the semiconductor partially resemble those of an intrinsic one.

In disordered samples, elastic scattering on impurities usually plays a considerable role at not-too-high temperatures (before the phonons take over).
Impurities break the Galilean invariance of the system and drastically enhance the  overall particle collision intensity~\cite{AltshulerAronovSpivak}.
One should be careful here and remember that at low temperatures, electron-electron repulsion can be complemented by their phonons-mediated attraction:
Even at very low temperatures, electrons might start forming Cooper pairs accompanied by the emission of virtual phonons.
Moreover, before the transition to the superconducting state, the superconducting fluctuations (short-lifetime Cooper pairs) start to play a role, leading to the paraconductivity~\cite{Varlamov} as an additional contribution to normal electron gas conductivity.


The interplay of particle scattering on impurities, Coulomb interaction, and photoinduced transport in general are especially interesting subjects regarding the emerging two-dimensional (2D) Dirac materials such as graphene~\cite{RefGraphene2005, PhysRevLett.99.236809} and transition-metal dichalcogenides (TMDs)~\cite{Mak1489, XiaoVHE, PhysRevB.77.235406}.
In addition to the regular translational momentum and spin degrees of freedom, these materials, possessing a hexagonal lattice, also host the valley degree of freedom.
It represents a quantum number describing the corners $K$ and $K^{\prime}$ of the Brillouin zone.
This nontrivial reciprocal lattice structure provides new opportunities to control particle transport~\cite{PhysRevLett.99.236809} studied within the direction of research referred to as valleytronics~\cite{PhysRevB.77.235406}.
The key property underlying all the valleytronic effects is the valley selection rule, which literally means that the (low-energy) electrons in each valley interact with either left or right circular polarization of external light, providing valley-selective interband transitions in the monolayer.

\begin{figure*}
\includegraphics[width=1.9\columnwidth]{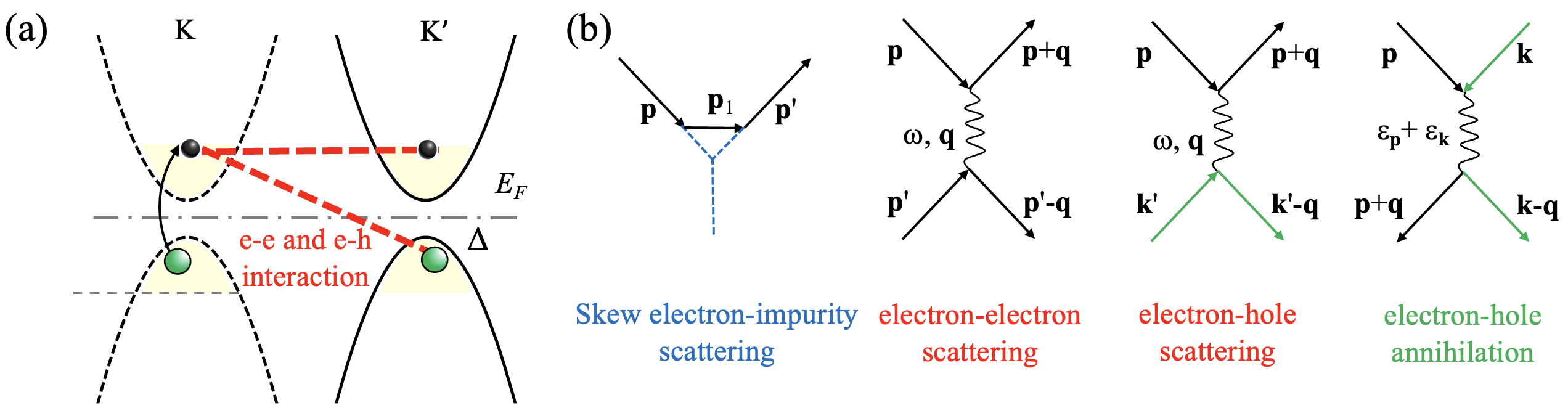}
\caption{(a) System schematic: The reciprocal space two-band structure of a 2D material. For a small-bandgap material with its Fermi level is in the bandgap (intrinsic semiconductor case), the electron and hole densities are distributed according to the Boltzmann statistics.
In the case of a wide-bandgap semiconductor, electrons and holes are excited due to the sample illumination by an external electromagnetic field.
In the degenerate electron gas regime, electrons occupy the conduction band up to the Fermi level (located in the conduction band).
(b) Quantum amplitude diagrams describing the skew electron-impurity scattering, electron-electron and electron-hole scattering and electron-hole annihilation processes, respectively.}
\label{Fig1}
\end{figure*}

Indeed, many of the Dirac materials represent direct-bandgap semiconductors (in the vicinity of the valleys $K$ and $K^\prime$) and within the optical range (for MoS$_2$ the gap is $1.66\,\mathrm{eV}$~\cite{XiaoVHE}).
It is a platform to study the valley photogalvanic~\cite{PhysRevB.99.075405, Entin2019, Entin2, PhysRevB.107.085415, PhysRevB.106.144502}, valley acoustoelectric, and the valley Hall effects~\cite{Glazov2DMat22, PhysRevB.109.085301, arXiv:2310.17738, PhysRevLett.122.256801, PhysRevB.103.035434, PhysRevB.100.121405, PhysRevB.102.235405,
PhysRevB.102.155302, Kovalev2018NJP, PhysRevLett.132.096302}.
The valley Hall effect has attracted particularly high attention in recent years.
It directly relates to the family of anomalous Hall effects: The transport phenomena, where the transverse current of charged carriers does not directly relate to the action of the Lorentz force~\cite{10.1080/14786448008626936, RevModPhys.82.1539, DyakonovBook}.
In the case of VHE, charge particles propagate in different transverse directions in different valleys.

Three specific mechanisms constitute the basis of VHE in nonmagnetic systems: (a) the asymmetric (skew) scattering, (b) the side jump, and (c) the anomalous velocity (or the Berry phase).
These mechanisms have recently been studied theoretically when applied to two-valley Dirac monolayers under different external drag forces: The static electric field, the phonon, and the photon drag~\cite{PhysRevB.102.155302}.
However, this theory is based on one-particle approximation, disregarding possible interparticle scattering processes.

In this Letter, we study the Coulomb interaction-renormalized VHE in gapped two-valley monolayer materials (such as gapped graphene, TMDs, etc.) in two regimes: (i) the $n$-doped monolayer (the high-density regime), when the electron gas is degenerate, and (ii) the low-density regime, when the monolayer is intrinsic, and both the electrons and holes take part in the transport phenomena and satisfy the Boltzmann statistics.
Case (ii) also covers the samples exposed to valley-selective light, as discussed below.
We will consider disordered samples at low temperatures when electron (and hole) impurity relaxation times are much smaller than the interparticle collision time.
In this regime, the interparticle collisions represent a correction to the bare VHE determined by the particle scattering on impurities~\cite{Lithuanian}.
Furthermore, we treat the anisotropic skew scattering of charge carriers on impurities as the principal mechanism of VHE~\cite{PhysRevB.102.155302}.

It should be noted that in very clean samples, another skew-scattering mechanism emerges: The interparticle skew scattering (particles skew-scatter on particles but not impurities).
We will disregard this mechanism as we consider disordered samples, thus focusing on the influence of direct particle-particle scattering on the disorder-induced VHE.
We will show that inter-particle corrections determine the temperature behavior of VHE at low temperatures.
It should also be noted that at low temperatures, the Cooper channel of interparticle scattering can contribute to the anomalous Hall effect~\cite{LI2020168137}; however, we will not consider this regime in this Letter.


{\it General framework of bare impurity-induced VHE.}
In the calculations, we employ the two-band spinless model with the parabolic dispersions of electrons and holes with equal effective masses (Fig.~\ref{Fig1}). Equilibrium densities of electrons and holes are determined either by the temperature or by the intra-valley pumping by means of an external electromagnetic (EM) field. The first case is realized in small bandgap materials, such as gapped graphene, whereas the second one takes place in wide-bandgap 2D materials, such as transition metal dichalcogenide monolayers, where the thermally-activated electron and hole densities are small due to the large value of the bandgap.
We consider the diffusive regime of the charge carriers' motion, assuming that the temperature is low enough such that the particle-impurity collision rate exceeds the particle-particle collision rate~\cite{Lithuanian}.
This approximation allows for using the Boltzmann equations and treating the particle-particle collision integral via successive approximations.

Let us first show that this approach provides known formulas for the bare VHE~\cite{PhysRevB.102.155302}
In the absence of interparticle collisions, the Boltzmann equations for electrons (momentum $\bf p$) and holes (momentum $\bf k$) can be written as
\begin{gather}
\label{bare1.1}
{\bf F}\cdot\frac{\partial f_{\bf k}}{\partial {\bf k}}
= Q^{s}_{hi}\{f_{\bf k}\}+Q^{a}_{hi}\{f_{\bf k}\},\\
\label{bare1.2}
-{\bf F}\cdot\frac{\partial f_{\bf p}}{\partial {\bf p}}
= Q^{s}_{ei}\{f_{\bf p}\}+Q^{a}_{ei}\{f_{\bf p}\},
\end{gather}
where ${\bf F}=e{\bf E}$ with $\mathbf{E}$ the electric field, $e>0$, and $Q^{s}$ and $Q^{a}$ are symmetric and asymmetric parts of electron- and hole-impurity collision integrals.
The former can be treated in the relaxation time approximation, $Q^{s}_{h(e)i}\{f_{{\bf k}({\bf p})}\}=-(f_{{\bf k}({\bf p})}-n_{{\bf k}({\bf p})})/\tau_i$, where $n_{{\bf k}({\bf p})}$ are the equilibrium hole (electron) distribution functions.
The anisotropic collision integral reads $Q^{a}_{ei}\{f_{\bf p}\}=\sum_{{\bf p}'}W_{{\bf p}'{\bf p}}f_{{\bf p}'}$, where the anisotropic scattering probability is~\cite{PhysRevB.102.155302} $W_{{\bf p}'{\bf p}}=\eta_eW_0[{\bf p}\times{\bf p}']_z\delta(\epsilon_{{\bf p}'}-\epsilon_{{\bf p}}),\,W_0=-2\pi u_0v^2/\tau_i\Delta^2$, with $u_0$ the strength of the short-range impurity potential, and $\Delta$ the monolayer bandgap, $v$ the  band parameter (in the framework of the two-band model), $\eta_e=\pm1$ the electron valley index, and finally, $\epsilon_{\textbf{p}}=\textbf{p}^2/2m$ the electron dispersion.
For the holes, the anisotropic scatting integral is determined by a similar expression, $W_{{\bf k}'{\bf k}}=\eta_hW_0[{\bf k}\times{\bf k}']_z\delta(\epsilon_{{\bf k}'}-\epsilon_{{\bf k}})$, with $\epsilon_{\bf k}=\textbf{k}^2/2m$ and $\eta_h$ the hole valley index (note, $\eta_e=-\eta_h$ for the same valley).

Solving Eqs.~\eqref{bare1.1} and \eqref{bare1.2} via successive approximations with respect to the anisotropic impurity scattering (assuming that $\delta f_{\textbf{p}(\textbf{k})}=f_{\textbf{p}(\textbf{k})}-n_{\textbf{p}(\textbf{k})}$ could be approximated as $\delta f_{\textbf{p}(\textbf{k})}=\delta f^{(0)}_{\textbf{p}(\textbf{k})}+\delta f^{(1)}_{\textbf{p}(\textbf{k})}$) gives  expressions for the zero-order correction to the distribution functions,
\begin{gather}\label{bare2}
\delta f^{(0)}_{\bf k}=-\tau_i({\bf F}\cdot{\bf v}_{\bf k})n'_{\bf k},\,\,\,\,
\delta f^{(0)}_{\bf p}=\tau_i({\bf F}\cdot{\bf v}_{\bf p})n'_{\bf p},
\end{gather}
where $n'_\mathbf{p}=\partial n_\mathbf{p}/\partial\epsilon_\mathbf{p}$, and for the first-order corrections with respect to anisotropic impurity scattering, $\delta f^{(1)}_{\bf k}=\tau_iQ^a_{he}\{\delta f^{(0)}_{\bf k}\}$ and $\delta f^{(1)}_{\bf p}=\tau_iQ^a_{he}\{\delta f^{(0)}_{\bf p}\}$:
\begin{gather}\label{bare3}
\delta f^{(1)}_{\bf k}=-\tau_i^2\eta_hW_0\sum_{{\bf k}'}
[{\bf k}\times{\bf k}']_z\delta(\epsilon_{{\bf k}'}-\epsilon_{{\bf k}})({\bf F}\cdot{\bf v}_{{\bf k}'})n'_{{\bf k}'}
\\
\nonumber
\delta f^{(1)}_{\bf p}=\tau_i^2\eta_eW_0\sum_{{\bf p}'}
[{\bf p}\times{\bf p}']_z\delta(\epsilon_{{\bf p}'}-\epsilon_{{\bf p}})({\bf F}\cdot{\bf v}_{{\bf p}'})n'_{{\bf p}'}.
\end{gather}

The total bare VHE electric current density reads ${\bf j}=e\sum_{{\bf k}}{\bf v}_{\bf k}\delta f^{(1)}_{\bf k}-e\sum_{{\bf p}}{\bf v}_{\bf p}\delta f^{(1)}_{\bf p}$. For electrons, in the case $\mathbf{F}=eE_x\hat{x}$, we find
\begin{eqnarray}
{j}_y^{e}&=&\sigma^{(0)}_{yx}E_x=
-2\eta_eW_0e^2\tau_i^2
\frac{m^2}{(2\pi)^2}
\langle\epsilon_{\textbf{p}}\rangle
E_x\int\limits_0^{\infty}d\epsilon_{\textbf{p}}n_{\textbf{p}},
\end{eqnarray}
where $\sigma^{(0)}_{yx}$ is the transverse conductivity of the bare VHE, and $\langle\epsilon_{\textbf{p}}\rangle=\sum_\mathbf{p}\epsilon_\mathbf{p}n_\mathbf{p}/\sum_\mathbf{p}n_\mathbf{p}=\int\limits_0^\infty d\epsilon_{\textbf{p}}\epsilon_{\textbf{p}}n_{\textbf{p}}/\int\limits_0^\infty d\epsilon_{\textbf{p}}n_{\textbf{p}}$ is the mean value of the particle kinetic energy.
Furthermore, for $\sigma^{(0)}_{yx}=\sigma^h_{yx}+\sigma^e_{yx}$ we find (restoring $\hbar$)
\begin{gather}
\label{bare4}
\sigma^{(0)}_{yx}=-2\sigma_D(\eta_h+\eta_e)
\left(\frac{W_0m^2}{2\pi\hbar}\right)
\frac{\langle\epsilon_{\mathbf{p}(\mathbf{k})}\rangle\tau_i}{\hbar},
\end{gather}
where $\sigma_D=e^2n_i\tau_i/m$ is the Drude conductivity, and $n_i$ is the
particle density.
For an intrinsic monolayer, when electrons and holes satisfy the Boltzmann statistics, $\langle\epsilon_\mathbf{p}\rangle=\langle\epsilon_\mathbf{k}\rangle=T$ with $T$ the temperature, and $n_i=\sqrt{n_en_h}$ with equal electron and hole densities, $n_e=n_h$.
Instead, for an n-doped monolayer (degenerate electron gas case), the contribution from the holes is absent, whereas $\langle\epsilon_\mathbf{p}\rangle=\mu/2$ with $\mu$ the Fermi energy, and $n_i\equiv n_e$, where $n_e$ is the concentration of degenerate electrons.
Formula~\eqref{bare4} coincides with a known result~\cite{PhysRevB.102.155302}.



{\it Electron-electron scattering.}
The Boltzmann equation, describing (i) the isotropic and anisotropic electron scattering on impurities and (ii) the electron-electron scattering reads
\begin{gather}\label{ee1}
-{\bf F}\cdot\frac{\partial f_{\bf p}}{\partial {\bf p}}
=
Q^{s}_{ei}\{f_{\bf p}\}
+Q^{a}_{ei}\{f_{\bf p}\}
+Q_{ee}\{f_{\bf p}\}.
\end{gather}
%
The first two terms on the right-hand side were defined above.
The last term describes the e-e scattering:
\begin{eqnarray}\label{ee2}
&&Q_{ee}\{f_{\bf p}\}=2\pi\sum_{{\bf p}',{\bf k}',{\bf k}}|U_{{\bf p}'-{\bf p}}|^2
\\
\nonumber
&&~~~\times[(1-f_{\bf k})(1-f_{\bf p})f_{{\bf k}'}f_{{\bf p}'}-(1-f_{{\bf k}'})(1-f_{{\bf p}'})f_{{\bf k}}f_{{\bf p}}]\\
\nonumber
&&~~~~~\times\delta(\epsilon_{{\bf k}'}+\epsilon_{{\bf p}'}-\epsilon_{{\bf k}}-\epsilon_{{\bf p}})\delta_{{\bf k}'+{\bf p}'-{\bf k}-{\bf p}}.
\end{eqnarray}
Here, $U_{{\bf p}'-{\bf p}}$ is a Fourier image of the e-e interaction potential.
In this section, both ${\bf p}$ and ${\bf k}$ correspond to electron momenta (in the subsequent sections, ${\bf k}$ stands for the hole momentum).

To find the conductivity, the solution of Eq.~\eqref{ee1} should be linearized with respect to the external force ${\bf F}$, thus $f_{\bf p}=n_{\bf p}+\delta f_{\bf p}$, where $\delta f_{\bf p}$ satisfies the linearized version of Eq.~\eqref{ee1}
\begin{gather}\label{ee3}
-({\bf F}\cdot{\bf v}_{\bf p})n'_{\bf p}
=-\frac{\delta f_{\bf p}}{\tau_i}
+Q^{a}_{ei}\{\delta f_{\bf p}\}
+Q_{ee}\{\delta f_{\bf p}\},
\end{gather}
where the linearized version of e-e collision integrals reads
\begin{widetext}
\begin{eqnarray}\label{ee4}
Q_{ee}\left\{\delta f_{\textbf{p}}\right\}&=&-2\pi\sum_{\mathbf{p}',\mathbf{k},\mathbf{k}'}
|U_{{\bf p}'-{\bf p}}|^2
\delta(\epsilon_{{\bf k}'}+\epsilon_{{\bf p}'}-\epsilon_{{\bf k}}-\epsilon_{{\bf p}})
\delta_{{\bf k}'+{\bf p}'-{\bf k}-{\bf p}}
\\
\nonumber
&&\times\Bigl[\delta f_{\bf p}[(1-n_{\bf k})n_{{\bf k}'}n_{{\bf p}'}+n_{\bf k}(1-n_{{\bf k}'})(1-n_{{\bf p}'})]
-\delta f_{{\bf p}'}[(1-n_{{\bf k}})(1-n_{{\bf p}})n_{{\bf k}'}+n_{\bf k}n_{\bf p}(1-n_{{\bf k}'})]\\
\nonumber
&&~~~+\delta f_{\bf k}[(1-n_{\bf p})n_{{\bf k}'}n_{{\bf p}'}+n_{\bf p}(1-n_{{\bf k}'})(1-n_{{\bf p}'})]
-\delta f_{{\bf k}'}[(1-n_{{\bf k}})(1-n_{{\bf p}})n_{{\bf p}'}+n_{\bf k}n_{\bf p}(1-n_{{\bf p}'})]\Bigr].
\end{eqnarray}
\end{widetext}

Since we are calculating the Coulomb corrections to the skew-scattered electrons, the conductivity must be determined by the product of weak scattering terms $\sim Q^a\cdot Q_{ee}$.
Such terms emerge in the second order corrections of the iteration procedure if we additionally assume $\delta f_{\textbf{p}}=\delta f^{(0)}_{\textbf{p}}+\delta f^{(1)}_{\textbf{p}}+\delta f^{(2)}_{\textbf{p}}$. Indeed,
the first-order correction reads $\delta f^{(1)}_{\bf p}=
\tau_iQ^{a}_{ei}\{\delta f^{(0)}_{\bf p}\}+\tau_iQ_{ee}\{\delta f^{(0)}_{\bf p}\}$, whereas the second-order corrections yield
$\delta f_{\bf p}^{(2)}=\tau_i(Q_{ei}^a\{\tau_iQ_{ee}\{\delta f_{\bf p}^{(0)}\}\}
+Q_{ee}\{\tau_iQ^a_{ei}\{\delta f_{\bf p}^{(0)}\}\})$, where we keep only the cross-terms $Q_{ee}\cdot Q^a$, as the terms proportional to $Q_{ee}^2,\,(Q^a_{ei})^2$ do not contain either the skew-scattering--related terms or the e-e--related ones.

An algebraic analysis (provided in the Supplemental Material~\cite{[{See Supplemental Material for the details of derivations at\\ }]SMBG}) gives expressions for the e-e corrections to VHE conductivity.
If the electron gas is degenerate, then (restoring $\hbar$)
\begin{eqnarray}
\nonumber
\sigma^{(d)}_{yx}&=&
\frac{2\pi}{3}
\sigma_D
(\eta_e-\eta_e')\left(\frac{W_0m^2}{2\pi\hbar}\right)
\left(\frac{T\tau_i}{\hbar}\right)^2
\left(\frac{e^2}{\hbar v_F\varepsilon}\right)^2\\
\label{EqMain01}
&&\times
\left[\ln\left(1+\frac{2p_F}{\hbar q_s}\right)-\frac{2p_F}{\hbar q_s+2p_F}\right],
\end{eqnarray}
where $\sigma_D=e^2n_e\tau_i/m$ is a Drude conductivity. Here, we used the screened Coulomb interaction $U_{q}=2\pi e^2/\varepsilon(q+q_s)$, where $q_s=me^2/\varepsilon\hbar^2$ is a screening wave vector
and $\varepsilon$ is a dielectric constant.

In the case of non-degenerate (Boltzmann) statistics of the electron gas, we can disregard the screening and use pure 2D Coulomb potential to find
\begin{gather}\label{EqMain02}
\sigma^{(nd)}_{yx}
=
\sigma_D
(\eta_e-\eta_e')
5\pi^2
\left(\frac{W_0m^2}{2\pi\hbar}\right)
\left(\frac{e^2\sqrt{n_e}}{\varepsilon T}\right)^2
\left(\frac{T\tau_i}{\hbar}\right)^2.
\end{gather}
The analysis of these two expressions is presented in the Discussion section below.


{\it Electron-hole scattering.}
In the case of an intrinsic semiconductor, both electrons and holes contribute to the Hall conductivity of the system.
Similarly to the e-e case, we start with the Boltzmann equations for electron and hole nonequilibrium distribution functions,
\begin{eqnarray}
\nonumber
-{\bf F}\cdot\frac{\partial f_{\bf p}}{\partial {\bf p}}
&=&
-\frac{f_{\bf p}-n_{\bf p}}{\tau_i}+Q^{a}_{ei}\{f_{\bf p}\}+Q_{eh}\{f_{\bf p},f_{\bf k}\},
\\
\nonumber
{\bf F}\cdot\frac{\partial f_{\bf k}}{\partial {\bf p}}
&=&
-\frac{f_{\bf k}-n_{\bf k}}{\tau_i}+Q^{a}_{hi}\{f_{\bf k}\}+Q_{he}\{f_{\bf p},f_{\bf k}\},
\end{eqnarray}
where
\begin{eqnarray}
\nonumber
Q_{ei}^a\left\{f_{\bf p}\right\}&=&-\eta_eW_0\frac{2\pi}{\tau_i}
\sum_{\mathbf{p}'}
[\mathbf{p}\times\mathbf{p}']_z f_{\mathbf{p}'}
\delta(\epsilon_{\textbf{p}}-\epsilon_{\textbf{p}'}),
\\
\nonumber
Q_{hi}^a\left\{f_{\bf k}\right\}&=&-\eta_h W_0\frac{2\pi}{\tau_i}
\sum_{\mathbf{k}'}
[\mathbf{k}\times\mathbf{k}']_z f_{\mathbf{k}'}
\delta(\epsilon_{\textbf{k}}-\epsilon_{\textbf{k}'}).
\end{eqnarray}
In this section, the electron momenta will be denoted by ${\bf p}$ and ${\bf p}'$, and the hole momenta by ${\bf k}$ and ${\bf k}'$; and we will not use the superscripts `e' and `h' in the distribution functions for brevity.

The linearized e-h collision integral reads
\begin{eqnarray}
\label{EqQeh1}
&&Q_{eh}\{\delta f_{\bf p},\delta f_{\bf k}\}=
-2\pi\sum_{\mathbf{p}',\mathbf{k},\mathbf{k}'}
(|U_{{\bf p}'-{\bf p}}|^2+|U_{\mathbf{p}+\mathbf{k}}|^2)\\
\nonumber
&&~~~\times
\left(
\delta f_{\bf p}n_{\bf k}
-\delta f_{\bf p'}n_{\bf k'}
+\delta f_{\bf k}n_{\bf p}
-\delta f_{\bf k'}n_{\bf p'}
\right)\\
\nonumber
&&~~~~~\times
\delta(\epsilon_{{\bf k}'}+\epsilon_{{\bf p}'}-\epsilon_{{\bf k}}-\epsilon_{{\bf p}})
\delta_{{\bf k}'+{\bf p}'-{\bf k}-{\bf p}},
\end{eqnarray}
and $Q_{he}\{\delta f_{\bf p},\delta f_{\bf k}\}$ acquires the same form.
Here, the first term $U_{\mathbf{p}-\mathbf{p}'}$ describes direct electron-hole scattering, whereas the second term $U_{\mathbf{p}+\mathbf{k}}$ stands for the electron-hole annihilation.
The latter can play an essential role in narrow gap materials such as gapped graphene~\cite{PhysRevB.109.085424} and, thus, must be accounted for.
The linear corrections satisfy the equations
\begin{eqnarray}
\frac{\delta f_{\bf p}}{\tau_i}={\bf F}\cdot\frac{\partial f_{\bf p}}{\partial {\bf p}}+Q_{ei}^a\{\delta f_{\bf p}\}+Q_{eh}\{\delta f_{\bf p},\delta f_{\bf k}\},
\\
\frac{\delta f_{\bf k}}{\tau_i}=-{\bf F}\cdot\frac{\partial f_{\bf k}}{\partial {\bf k}}+Q_{ei}^a\{\delta f_{\bf k}\}+Q_{he}\{\delta f_{\bf p},\delta f_{\bf k}\}.
\end{eqnarray}
We solve these equations by iterations by expanding $\delta f_{\textbf{p}(\textbf{k})}=\delta f^{(0)}_{\textbf{p}(\textbf{k})}+\delta f^{(1)}_{\textbf{p}(\textbf{k})}+\delta f^{(2)}_{\textbf{p}(\textbf{k})}+...$ such that
$\delta f_{\bf p}^{(0)}=\tau_i(\mathbf{F}\cdot\mathbf{v}_{\mathbf{p}})n_\mathbf{p}'$,
$\delta f_{\bf k}^{(0)}=-\tau_i(\mathbf{F}\cdot\mathbf{v}_\mathbf{k})n_\mathbf{k}'
$, and
$\delta f_{\bf p}^{(1)}=\tau_i(Q_{ei}^a\{\delta f_{\bf p}^{(0)}\}+Q_{eh}\{\delta f_{\bf p}^{(0)},\delta f_{\bf k}^{(0)}\})$, $
\delta f_{\bf k}^{(1)}=\tau_i(Q_{ei}^a\{\delta f_{\bf k}^{(0)}\}+Q_{he}\{\delta f_{\bf p}^{(0)},\delta f_{\bf k}^{(0)}\})$, and so on.

Performing the algebraic derivations, the Hall conductivity yields~\cite{SMBG}
\begin{eqnarray}
\label{Eqeh01}
\sigma^{(eh)}_{yx}&=&-W_0e^2\tau_i^3
\sum_{\mathbf{p},\mathbf{p}',\mathbf{k}',\mathbf{k}}
\left(|U_{\mathbf{p}-\mathbf{p}'}|^2+|U_{\mathbf{p}+\mathbf{k}}|^2\right)\\
\nonumber
&&\times
(k_y-p_y)
\left[\eta_e(p_y\epsilon_{\textbf{p}}-p_y'\epsilon_{\textbf{p}'})
-
\eta_h(k_y\epsilon_{\textbf{k}}-k_y'\epsilon_{\textbf{k}'})\right]\\
\nonumber
&&\times(n_{\bf p}-n_{{\bf p}'})
(n_{\bf k}-n_{{\bf k}'})\delta(\mathbf{p}+\mathbf{k}-\mathbf{p}'-\mathbf{k}')\\
\nonumber
&&\times
\int d\omega \frac{dN_\omega}{d\omega}
\delta(\epsilon_{\mathbf{k}'}-\epsilon_\mathbf{k}-\omega)
\delta(\epsilon_{\mathbf{p}'}-\epsilon_\mathbf{p}+\omega).
\end{eqnarray}
%
Furthermore, it is convenient to split Eq.~\eqref{Eqeh01} into two terms, $\sigma^{(eh)}_{yx}=\sigma^{(h1)}_{yx}+\sigma^{(h2)}_{xy}$, where $\sigma^{(h1)}_{yx}$ describes e-h scattering and $\sigma^{(h2)}_{yx}$ corresponds to e-h annihilation.
After algebraic calculations~\cite{SMBG}, we find
\begin{gather}
\label{EqMain03}
\sigma^{(h1)}_{yx}
=
\sigma_D
(\eta_e+\eta_h)
5\pi^2
\left(\frac{W_0m^2}{2\pi\hbar}\right)
\left(\frac{e^2\sqrt{n_i}}{\varepsilon T}\right)^2\left(\frac{T\tau_i}{\hbar}\right)^2,
\end{gather}
and
\begin{gather}
\label{EqMain04}
\sigma^{(h2)}_{yx}
=
\sigma^{(h1)}_{yx}
\frac{2}{5}
\int\limits_0^\infty\frac{x(x^2+2)e^{-x^2/2}}{(x+\frac{\hbar q_T}{p_T})^2}dx,
\end{gather}
where in the last expression, we took into account the static screening for the non-degenerate statistics of electrons and holes
$U_q=2\pi e^2/\varepsilon(q+q_T)$ with $q_T=2\pi e^2(2n_i)/T$ the screening wave vector,
and $p_T^2=2mT$. In Eq.\eqref{EqMain03} we assume that electron and hole densities are equal, $n_i=n_e=n_h$.


{\it Results and discussion.}
Formulas~\eqref{EqMain01},~\eqref{EqMain02},~\eqref{EqMain03} and~\eqref{EqMain04} represent the central results of this Letter.
Let us analyze them separately and compare them.

For both the cases of degenerate and non-degenerate electron gases (Eqs.~\eqref{EqMain01} and~\eqref{EqMain02}), the corrections are proportional to $(\eta_e-\eta_e')$, which means that electrons from valley $K$ scatter with electrons from $K^\prime$, and the scattering of electrons from the same valley is suppressed.
The same argument is valid for the electron-hole scattering: In Eqs.~\eqref{EqMain03} and~\eqref{EqMain04}, the term $(\eta_e+\eta_h)$ is only nonzero if the particles belong to different valleys (since $\eta_h=-\eta_e$).

Furthermore, Eqs.~\eqref{EqMain02} and~\eqref{EqMain03} resemble each other up to the valley index-dependent term, as expected: e-e and e-h-mediated scattering should be similar for an intrinsic (or nearly intrinsic) semiconductor with equal effective masses of the carriers of charge.

From Eq.~\eqref{bare4}, we extract that for a degenerate electron gas, the bare VHE conductivity reads as
\begin{eqnarray}
\label{EqSig0d}
\sigma^{(0,d)}_{yx}=-2\sigma_D\eta_e
\left(\frac{W_0m^2}{2\pi\hbar}\right)
\frac{\mu\tau_i}{2\hbar},
\end{eqnarray}
and for a nondegenerate gas,
\begin{eqnarray}
\label{EqSig0nd}
\sigma^{(0,nd)}_{yx}=-2\sigma_D\eta_e
\left(\frac{W_0m^2}{2\pi\hbar}\right)
\frac{T\tau_i}{\hbar}.
\end{eqnarray}
The Coulomb interaction-induced correction~\eqref{EqMain01} in the case of a degenerate electron gas depends on the temperature ($\sigma_{yx}^{(d)}\sim T^2$) in contrast to the bare degenerate electron gas VHE if in Eq.~\eqref{EqSig0d} we disregard the dependence of the chemical potential on temperature (which is extremely weak for a degenerate gas).

Instead, in the case of the non-degenerate electrons, the correction~\eqref{EqMain02} might not depend on temperature, as compared with the bare VHE~\eqref{EqSig0nd} (which is linear in $T$).
Indeed, the explicit dependence on temperature disappears in~\eqref{EqMain02} ($T^2$ cancels out in the numerator and denominator there).
However, $n_e$ in~\eqref{EqMain02} in the case of intrinsic material at thermal equilibrium might be temperature dependent. This case is essential for relatively narrow-bandgap materials, such as a gapped graphene. In the case of wide-bandgap materials, such as MoS$_2$ in the ``intrinsic'' regime, the thermally activated electron and hole densities are very small.
In this case, a nonzero VHE current can be achieved by destroying the time-reversal symmetry by, e.g., the illumination of the sample by a circularly polarized light with the frequency of the order of the interband transition energy, thus populating one of the valleys~\cite{PhysRevB.77.235406} and keeping the other valley at equilibrium.
Under optical excitation, $n_e$ and $n_h$ are temperature-independent since these densities are controlled by an external pump.

Then, for wide-bandgap materials, the correction due to e-h scattering~\eqref{EqMain03} in the case of the non-degenerate electrons and holes, is also temperature-independent.
However, the e-h annihilation~\eqref{EqMain04} does reveal a peculiar temperature dependence due to the $q_T$-dependent term in the integral.
Let us elaborate on that.
\begin{figure}
\includegraphics[width=0.9\columnwidth]{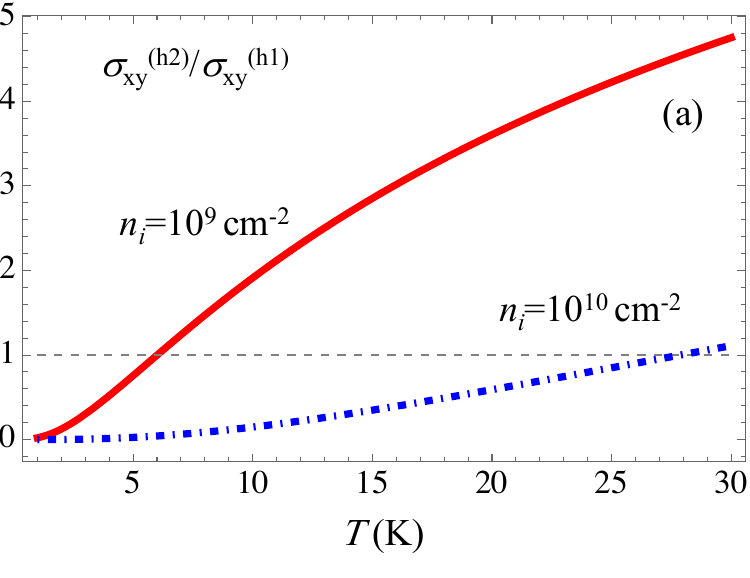}
\includegraphics[width=0.9\columnwidth]{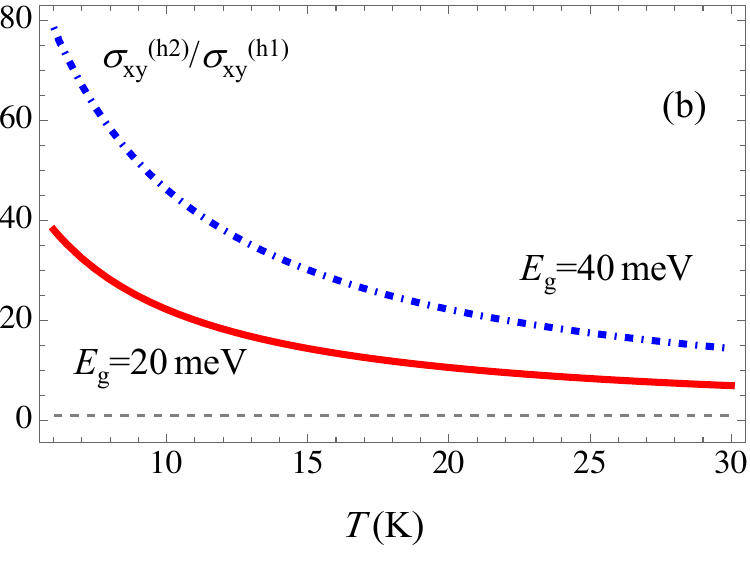}
\caption{The ratio of e-h exchange (annihilation) and direct e-h scattering-mediated conductivities in the case of (a) MoS$_2$, where we used the temperature-independent effective intrinsic density of particles $n_i$ (two values), and (b) gapped graphene with the temperature-dependent $n_i$ for two values for the bandgap $E_g$.
In (b), the annihilation processes are dominant over the whole range of reasonable temperatures.}
\label{Fig2}
\end{figure}
Figure~\ref{Fig2}(a) shows the ratio $\sigma^{(h2)}_{xy}/\sigma^{(h1)}_{xy}$ for an intrinsic MoS$_2$ (an example of wide-bandgap material) in the presence of an additional illumination of one of the valleys.
The curve starts from zero and crosses unity at some temperature.
Thus, we conclude that electron-hole annihilation grows with temperature and can eventually become the predominant mechanism of the Coulomb-mediated VHE.

The situation is different in gapped graphene (an example of small-bandgap material).
Figure~\ref{Fig2}(b) shows the ratio $\sigma^{(h2)}_{xy}/\sigma^{(h1)}_{xy}$, where nonzero $n_i$ is due to the finite temperature, $n_i=n_e=n_h=(mT/2\pi\hbar^2)\exp(-E_g/2T)$, and $m_g=E_g/2v_0^2$ with $v_0$ the graphene Fermi velocity.
Here, the processes of e-h annihilation are always dominant (up to very high temperatures, where phonon-mediated processes become predominant).

Finally, let us compare the Coulomb scattering-mediated effect and the bare VHE.
If we divide the nondegenerate electron gas Hall conductivity Eq.~\eqref{EqMain02} by the bare conductivity Eq.~\eqref{EqSig0nd}, we acquire the factor $T\tau_i\gg1$ among other.
It represents the ratio of the particles' kinetic energy and the impurity-mediated broadening $\hbar/\tau_i$.
This factor can compensate for the other small factors in the ratio $\sigma_{xy}^{(nd)}/\sigma_{xy}^{(0,nd)}$, and it might even result in the dominance of the Coulomb terms. It should be noted that the considered effects take place in monolayer materials with the broken inversion symmetry of crystal lattice. Such a broken symmetry can lead also to many features of elementary interparticle interactions (electron-electron interaction, electron-phonon interaction, etc) in various nanostructures (see, e.g., Refs.~[\onlinecite{Kibis-1992,Kibis-1999,Kibis-2001,Kibis-2002}]).


{\it Conclusions.}
We developed a theory of a Coulomb interaction-related contribution to the valley Hall effect and drew the following conclusions.
First, in the case of a degenerate electron gas in a semiconductor, the scattering of electrons from different valleys plays the major role in Coulomb interaction-induced valley Hall effect.
The Coulomb interaction--mediated contribution to the Hall conductivity depends on the temperature as $T^2$.
Second, in the case of an intrinsic semiconductor embedding non-degenerate electron and hole gases, the direct Coulomb scattering of particles from different valleys plays the major role for the effect in question up to some temperatures.
Above a certain temperature, electron-hole annihilation might become the dominating mechanism, if this temperature is lower than the Debye temperature, when phonons are expected to take over.
The Coulomb interaction-mediated contribution to the Hall conductivity does not depend on the temperature in the case of photogenerated non-degenerate electron and hole gases.

{\it Acknowledgments.}
This work was supported by the Institute for Basic Science in Korea (Project No. IBS-R024-D1), Ministry of Science and Higher Education of the Russian Federation (Project No. FSUN-2023-0006), and the Foundation for the Advancement of Theoretical Physics and Mathematics ``BASIS''.





\bibliography{biblio}
\bibliographystyle{apsrev4-2}


\end{document}